%% file: main.tex
\documentclass[journal]{IEEEtran}

\usepackage{xspace} 
\usepackage{amsfonts}
\usepackage{mathtools}
\usepackage{breqn}
\usepackage{float}

\usepackage{bbm}
\usepackage{epsfig,subfigure}
\usepackage{amssymb,amsmath}
\usepackage{enumerate}
\usepackage{latexsym}
\usepackage{booktabs}
\usepackage{array} 
\usepackage{comment}
\usepackage{todonotes}
\usepackage{url}
\usepackage{algorithm}
\usepackage{algorithmic}
\usepackage{caption}

\usepackage{graphicx}
\usepackage{multirow} 

\usepackage{amsthm}

\newtheorem{theorem}{Theorem}[section]
\newtheorem{lemma}[theorem]{Lemma}
\newtheorem{defn}{Definition}

\begin{document}



\title{Differential Privacy in Aggregated Mobility Networks: Balancing Privacy and Utility}




\author{Ammar~Haydari, ~\IEEEmembership{Student~Member,~IEEE,}, ~Chen-Nee~Chuah,~\IEEEmembership{Fellow,~IEEE}, Michael~Zhang,  Jane~Macfarlane, Sean~Peisert,~\IEEEmembership{Senior Member,~IEEE}

\thanks{Ammar Haydari and Chen-Nee Chuah are with the Department
of Electrical and Computer Engineering, University of California, Davis,
CA, 95616 USA (e-mail:ahaydari@ucdavis.edu; chuah@ucdavis.edu).}

\thanks{Sean Peisert is with Lawrence Berkeley National Laboratory, Berkeley, CA 94550 USA (e-mail: sppeisert@lbl.gov).}

\thanks{Jane Macfarlane is with the University of California, Berkeley, CA 94550 USA (e-mail: janemacfarlane@berkeley.edu).}

\thanks{Michael Zhang is with the Department of Civil and Environmental Engineering, University of California, Davis, CA 95616 USA (e-mail: hmzhang@ucdavis.edu).}
}

\maketitle

\begin{abstract}
 Location data is collected from users continuously to understand their mobility patterns. Releasing the user trajectories may compromise user privacy. Therefore, the general practice is to release aggregated location datasets. However, private information may still be inferred from an aggregated version of location trajectories. Differential privacy (DP) protects the query output against inference attacks regardless of background knowledge. This paper presents a differential privacy-based privacy model that protects the user's origins and destinations from being inferred from aggregated mobility datasets. \textcolor{black}{ This is achieved by injecting Planar Laplace noise to the user origin and destination GPS points. The noisy GPS points are then transformed into a link representation using a link-matching algorithm. Finally, the link trajectories form an aggregated mobility network. The injected noise level is selected using the Sparse Vector Mechanism. This DP selection mechanism considers the link density of the location and the functional category of the localized links. Compared to the different baseline models, including a k-anonymity method, our differential privacy-based aggregation model offers query responses that are close to the raw data in terms of aggregate statistics at both the network and trajectory-levels with maximum $9\%$ deviation from the baseline in terms of network length.} 
\end{abstract}





\input{introduction}
\input{Relatedwork.tex}

\input{Problem_Definition}

\input{Noise.tex}

\input{Results_new}
\input{Discussion.tex}
\input{Conclusion.tex}

\bibliographystyle{IEEEtran}
\bibliography{main.bib}

\end{document}

%% file: introduction.tex
 \section{Introduction}

With the proliferation of smartphones, GPS devices, and connected vehicles, geospatial-temporal datasets that capture the movements of these devices have emerged. The trajectories collected from these devices are excellent sources of mobility information for city planners and researchers. However, embedded in these datasets are details of the device owner's mobility patterns. Privacy concerns associated with lifestyle patterns, such as locations of home, work, and user specific points of interest, have inhibited the release of these datasets by the organizations collecting the data. As such, the datasets have remained siloed and their use has been limited to a few organizations. 

\textcolor{black}{In recent discussions between the California Public Utility Commission (CPUC) \cite{CPUC} and Transportation Network Companies like Uber and Lyft, deliberations have revolved around the anonymization and reporting of passenger trip data \cite{CPUC_decision}. The urgency to address these concerns stems from the Privacy, Security, and Transparency Committee of the Open Mobility Foundation \cite{OMF}, which is dedicated to ensuring the secure handling of mobility data. This demonstrates the need for innovations in mobility data protection techniques that balance privacy and utility, which has the potential to significantly influence policy decisions among multiple stakeholders, including data providers, neutral data-hosting third parties, and state governments.}

Typical approaches for dealing with privacy issues, such as k-anonymity or other privacy methods, often result in high utility loss for the type of transportation studies cities need. A naive approach of applying noise injection such as \cite{andres2013geo} to the individual GPS points can perturb the trajectory too much, particularly in urban areas where moving GPS points a few blocks away can dramatically change the path (e.g., due to one-way streets or locations of highway exits). 

Another approach to dealing with privacy issues is aggregating the data before releasing the dataset. However, \cite{xu2017trajectory} has shown that even aggregated outputs may still have privacy issues. The authors were able to re-identify user trajectories from the aggregated trajectories. Additional studies found that examining trajectories that were often repeated indicated unique mobility patterns that could be associated to an individual. Other efforts to privatize mobility data \cite{qardaji2013differentially, zhao2020novel}, resulted in a high utility loss.

Differential privacy (DP) is a statistical privacy-preserving technique \cite{dwork2006differential} that is designed to minimize leakage of information about individuals, while still preserving the characteristic patterns in the data. Differential privacy controls the degree of information that can potentially be exposed. The DP control parameters can be tuned empirically for specific datasets, and specific application use-cases \cite{Google-DP, Near2018Differential-Pr, Diffprivlib}. The goal of DP is to ensure that an adversary with background knowledge about the dataset cannot extract private information from the dataset. 

\textcolor{black}{Merely adding fixed noise to GPS data is not sufficient to protect location privacy. While the noise level can be enough for a dense areas, different level of noise might needed in sparsely populated regions. Our objective is to create a differential privacy (DP)-based method that ensures location privacy while retaining essential data for deriving transportation metrics. Our approach involves a network-aware noise injection algorithm (Laplace noise) using a sparse vector technique for adaptive noise range selection. By leveraging geospatial constraints, this model effectively anonymizes sensitive information.} 

\textcolor{black}{This work achieved promising results applying adaptive noise injection to origin destinations conditioned on the density of road segments to protect the privacy of individuals named \textit{a differentially private adaptive noise injection (DP-ANI) model} that generates an aggregated mobility network from raw GPS trajectory data.}

The contributions of this paper include:
\begin{itemize}
\item We propose a differential privacy-based noise injection (DP-ANI) model that perturbs the origin-destination GPS points in an adaptive manner based on the road network's density.

\item We apply the Sparce Vector Technique to select the adaptive range parameters privately.

\item We evaluate the impact of the degree of perturbation of the noise injection model by comparing the geospatial statistics derived from the released mobility network after applying our DP-ANI model compared to the raw mobility data. 
\end{itemize}

The remainder of the paper is organized as follows. We present related work in Section \ref{s:related}, an overview of the differential privacy in Section \ref{s:DP}, and our metrics and models in Section \ref{s:metrics}. We then present our differentially private adaptive noise injection (DP-ANI) model in Section \ref{s:Noise}. After evaluating the experimental results of our privacy model in Section \ref{s:results}, we discuss the limitations of this work in Section \ref{s:discuss}. Finally, Section \ref{s:conclusion} concludes the paper. 

%% file: Relatedwork.tex
\section{Background and Related Work}\label{s:related}

\subsection{The Importance of Privacy in Geospatial Datasets}
Geospatial mobility datasets describe movement of vehicles, bikes, scooters, or pedestrians, and are often collected from users with their permission. However, when these datasets are shared with third parties, significant privacy issues may arise. Removing individual identifiers is not enough to achieve strong privacy because it is well known that linkage attacks, using multiple datasets, can allow attackers to identify the users even if directly identifiers, such as name, social security number, etc... are excluded from the dataset \cite{nissim2017differential}. 
 
User patterns such as daily routines, extracurricular activities, and business activities are a part of behavioral privacy \cite{finn2013seven}. Mobility datasets capture these movements and have the potential to reveal information about behaviors of individuals and groups. Ideally, a privatized mobility dataset should maintain the overall movement patterns of the general population, while preventing the identification of small group or individual behaviors using inference attacks.

\subsection{Related Work on Location Privacy}
Protecting individual user location with DP has been studied extensively \cite{errounda2020analysis}. We can divide the location privacy models into two categories: privacy of online locations that shares the location data instantly and offline locations that uses for location dataset for extracting information. Privacy research on online location deals with the instant location information collected by user mobility applications such as navigation devices in real-time \cite{wang2018privacy}. Offline location privacy research focuses on movement data (time ordered locations) and aggregated mobility datasets.

\textbf{Privacy of Online Locations:} There are several DP-based location models in the literature for preserving the privacy of online locations where location information is protected before it reaches to the data-center. For example, location-based social networks provide privacy for every location sample \cite{wang2018privacy}.  

Generalizing and perturbing the actual location are the two popular approaches for location privacy. One of the early and well known location privacy techniques using DP models is called geo-indistinguishability \cite{andres2013geo} where the authors injected planar Laplace noise to the GPS points for hiding individual locations. A related study for real-time location sharing applications is proposed by \cite{elsalamouny2016differential} using circular noise functions. There are several privacy applications in the transportation domain inspired by the geo-indistinguishability \cite{shi2019deep, zhou2018achieving} for location privacy.

Another approach for location privacy is snapping individual points to a grid with lower resolution than the source data. The shared data is not the precise user location but is in a similar area. The authors presented a differentially private grid partitioning model for hiding user locations \cite{qardaji2013differentially}. Differential privacy for grid partitioning of spatial crowdsourcing applications is studied with an adaptive multi-level grid decomposition technique in \cite{wei2019differential}.

Anonymity-based privacy approaches seek to provide indistinguishability between users. Several k-anonymity privacy schemes in the case of real-time location sharing are presented in \cite{gedik2007protecting}, and \cite{fei2017k}. In \cite{ngo2015location}, authors presented a location privacy approach for online locations using cloaking area \cite{kalnis2007preventing} and differential privacy models. 

While there are many privacy models for online location sharing platforms, real-time trajectory sharing requires more sophisticated and advanced privacy models due to spatial and temporal correlations of location traces. There are several privacy protection attempts for the online trajectories \cite{chatzikokolakis2014predictive, xiao2017loclok}. The authors, in \cite{xiao2015protecting}, presented a DP-based planar isotropic location perturbation mechanism using a geometric sensitivity model.

\textbf{Privacy of Offline Locations:} Prefix-tree based aggregation is popular for achieving location privacy. One of the earlier methods creates a prefix tree and adds noise to the output node counts to achieve generalized trajectories \cite{chen2012differentially}. Another prefix-tree and DP-based trajectory privacy method is proposed in \cite{zhao2020novel} where authors used minimum description length method for clustering the trajectory segments on prefix-tree and injected a controlled noise to the count of clusters. Since generating differentially private trajectories requires a dense trajectory population to generalize or group trajectories, it is difficult to privatize individual trajectories. The methods presented in \cite{jiang2013publishing, shao2013publishing} aim to preserve the privacy of a trajectory using differential privacy. 

There are different research directions regarding the type of aggregated geospatial data that is released: trajectories, networks, and statistics. Recently, a trajectory aggregation mechanism with DP was presented in \cite{ghazi2022private}, where the aggregation range was found privately using the Sparse Vector Technique (SVT). Inspired by this paper, we apply the SVT to our private range mechanism. An aggregate mobility data publication approach using a count-min-sketch method is presented in \cite{zhang2021differential} for mobility distributions. This work evaluates the proposed aggregation mechanism against trajectory recovery attack model \cite{xu2017trajectory}. The framework studied in \cite{fan2013adaptive} applies filtering and adaptive sampling methods with differential privacy for sharing aggregate time-series statistics. The amount of aggregation is calculated with the proposed approach adaptively. Another aggregate geospatial statistic publishing approach with DP is presented in \cite{kellaris2014differentially} where a sliding-window methodology captures event consequences on data stream. A DP-based data aggregation model is presented in \cite{mir2013dp} using a dataset from user call records. The traces are classified into several pre-defined groups, and noise is injected to the count of those groups. An online data aggregation method with location privacy is studied in \cite{wang2016rescuedp} where authors presented a framework that dynamically groups and perturbs the statistics of aggregated locations using Laplace noise for achieving privacy protection. Markov decision modeling has recently been applied to aggregate mobility data privacy considering practical adversarial attacks to data privacy \cite{zhang2022privacy}.

Recently, several private companies made location privacy tools available on their platforms. Gratel Labs \cite{grateAI2020} offers a synthetic location sampling model using generative models. Here Technologies \cite{here2021} offers various location privacy techniques for protecting individual privacy, however, provenance of the data and the solutions are unknown. 

Other location privacy methods include hidden Markov models and k-anonymity. One of the offline trajectory privatization methods proposed in \cite{ou2016multi} used a hidden Markov model for protecting trajectory datasets against multi-user correlation attacks. An example of the k-anonymity trajectory privatization method is presented in \cite{monreale2010movement} where authors generalized movements to groups using a prefix-tree where leaves of the trees are removed if they have a value less than k. Transport network sharing for fleet vehicles is presented with k-anonymity and information theoretic approach using simulated ride-share datasets in \cite{he2019optimal} where the goal is to preserve the privacy of fleet trajectories by hiding the pickup and drop-off locations.    

\textcolor{black}{Several re-identification attacks were able identify individuals from publicly available datasets, such as NYC \cite{douriez2016anonymizing} and London bike sharing \cite{siddle2014}. The authors in \cite{tu2018new} experimented with an attack strategy that recovers trajectories from aggregated location datasets. The key assumption in the paper is that the aggregated data is not anonymized. However, our focus remains primarily theoretical in proposing a differential privacy (DP)-based model for protecting location privacy within aggregated mobility networks, providing theoretical assurances. The practical implications of these attacks on mobility networks are beyond the scope of our present work. Our immediate focus does not involve the practical privacy analysis of DP-ANI. Instead, we earmark this as a subject for future investigation and enhancement.}

\textcolor{black}{In contrast to the existing studies that concentrate on protecting individual data points or entities within a dataset, our proposed DP-ANI model considers the connectivity and spatial density of road segments. The road network-based density quantification ensures a more comprehensive privacy protection strategy for aggregated mobility dataset. The incorporation of both noise injection and private selection mechanisms within DP-ANI further strengthens its ability to protect sensitive information effectively. This emphasis on road network characteristics not only maintains the privacy guarantees but also enhances the applicability and reliability of our proposed DP-based approach in aggregated mobility datasets, establishing a new standard in location privacy.}

\subsection{Aggregation of Mobility Data}
Modeling mobility in urban regions often involves two main concepts: travel demand and infrastructure loading. Travel demand describes the mobility needs of a user population over a period of time. Infrastructure loading refers to the loading that the road network experiences as a function of the travel demand. The goal of transportation planning and operations is to ensure travel demand is served in the most efficient and safe manner. 

Aggregation is a common approach for managing privacy in transportation datasets \cite{popa2011privacy}. Individuals are clustered into an aggregated group of users that reduce the size and complexity of data.  An example of publicly-available, aggregated mobility data is the Uber Movement website \cite{Uber-Movement} where aggregated travel patterns for specified cities are released. The aggregation is done at both the temporal (one hour bins) and geospatial level (traffic analysis zone - TAZ).

\subsection{Differentially Private Movement Datasets}

Our mechanism preserves user-level privacy by hiding individual user participation in the aggregated dataset and prevents trajectory re-identification attacks by perturbing the origin and destination points. The amount of information loss associated with the applied privacy model is directly related to the granularity of the transportation problem of interest. In this paper, we focus on two transportation problems that require coarse or medium-level granularity and can be solved with our aggregated, differentially-private mobility network.

\textbf{Congestion Analysis:} An estimate of expected traffic congestion and the associated congestion mitigation plans are two key concerns for city planners. The transportation network is composed of links that define the road network. A zone refers to a certain area of the city and a collection of links. The capacity of links and their temporal changes are used to estimate traffic congestion using link-level aggregated statistics. Our privacy model generates differentially private aggregated link-level road metrics. 

\textbf{Major route identification:} Differentially private traffic network generated from our model is a directed graph which shows the origins and destinations of traffic flows at the link level. It is possible to identify major traffic routes that may result in congestion in selected links. This output also allows predicting future traffic behaviors using some data-driven prediction models. One could train a machine learning agent with the privatized query response to predict travel behavior for the next day. Authors in \cite{raadsen2020aggregation} studied major route and busy traffic with an aggregated bicycle dataset. Our mechanism perturbs the origin/destination of trajectories before aggregation using the density and attributes of the localized links and preserves the main traffic routes.

%% file: Problem_Definition.tex
\section{Differential Privacy: Overview}\label{s:DP}

Location privacy is the notion of privacy for aggregated mobility datasets. We require that the output of a query statistically guarantees privacy of individual user locations independent of the background knowledge. Differential privacy (DP) \cite{dwork2008differential} guarantees that modifying the single input value has a negligible effect on the output statistical query. In this section, we summarize the general definitions and metrics of DP that are applicable to our problem.

We introduce the privacy concerning data $\mathbf{X} \in \mathbf{\mathcal{X}}$ as vehicular mobility information in query $q \in {\mathcal{Q}}$. The data holder wants a mechanism that hides the sensitive information and reports the privacy preserved version of sensitive information using a randomized algorithm $\mathcal{A}: \mathbf{X} \times \mathcal{Q} \rightarrow \mathcal{D}$ where $ \mathcal{Q}$ is the query space and $\mathcal{D}$ is the output space. DP promises that the algorithm $\mathcal{A}$ is differentially private such that participation or removal of a record results in minimal changes to the output of a query.

Let us first define the neighboring datasets:
\begin{defn}\label{def:dist-one-element} [Neighboring Dataset] Considering two databases $\mathbf{X}$ and $\mathbf{X}'$, if they differ by only one element $\mathbf{x_i} \rightarrow \mathbf{x_i}'$ corresponding to a link on the network, they are neighboring datasets.
\end{defn}

The above definition formalizes the adjacent or neighboring dataset that plays a crucial role in differential privacy. 

\begin{defn} \label{def:DP} [$\epsilon$-Differential Privacy] Given  for every neighboring sets ${d}\subset {\mathcal{D}}$, a randomized algorithm $\mathcal{A}$ is $\epsilon$-differentially private if
\begin{equation}
    \Pr(\mathcal{A}(X) \in d) \leq e^{\epsilon}\Pr(\mathcal{A}(X') \in d)
\end{equation}
where $\epsilon$ is a positive real number and probability comes from the randomness of the algorithm. $\frac{\Pr(\mathcal{A}(X) \in d)}{\Pr(\mathcal{A}(X') \in d)}$ is the privacy leakage risk for the randomized algorithm $\mathcal{A}$.
\end{defn}

$\epsilon$-differential privacy is known as randomized response where adding or removing a single element from the database results in a similar probability. 
The smaller value of $\epsilon$ represents higher privacy guarantee and provides in-distinguishability. 

In differential privacy, the appropriate epsilon is typically determined based on the sensitivity of the underlying data. The definition of sensitivity is given in \cite{dwork2006differential} as follows:

\begin{defn} \label{def:sensitivity} [Sensitivity] For any query function $f$: $D \rightarrow R^{n}$, the sensitivity of $f$ is 
\begin{equation}
    \Delta f = \max_{\mathbf{X}, \mathbf{X}'} \left\|f(\mathbf{X})-f(\mathbf{X}')\right\|_{1}  
\end{equation}
for all datasets $\mathbf{X}$ and $\mathbf{X}'$.
\end{defn}

Input perturbation and output perturbation are the two ways to implement differential privacy. When we want to release an aggregated mobility network, one way to protect privacy is through input perturbation. Laplace mechanism adds a random noise sampled from Laplace distribution:
\begin{defn} \label{def:Laplace} [Laplace Mechanism] For any function $f$: $D \rightarrow R^{n}$, the mechanism  $\mathcal{A}$ gives $\epsilon$-DP as follows:
\begin{equation}
    \mathcal{A}(D)=f(D)+Laplace(\epsilon, R)
\end{equation}
\end{defn}

Noise injection to the input can be done with different noise functions depending on the application requirements. Section \ref{s:noise_sampling} describes our additive noise method in detail. 
\textcolor{black}{
For a given sequence of queries and a threshold $T$, a mechanism called sparse vector technique (SVT) outputs a vector indicating whether each query answer is above or below T. The goal is to find the first index from query output that is above the threshold. }

\begin{defn} \label{def:svt} [Sparse Vector Technique] \textcolor{black}{Suppose $f_1, f_2,..,f_k$:$\mathbf{X}\rightarrow R$ be set of functions and $T$ be a threshold for a database $\mathbf{X}$, the algorithm outputs binary outcome for each query answer $a_{i}$ if it is above the noisy threshold or not: $a_{i} \in \{\top, \bot\}^{k}$.}
\end{defn}

\begin{defn} \label{def:at} [AboveThreshold] \textcolor{black}{Given $f_1, f_2,..,f_k$:$\mathbf{X}\rightarrow R$ set functions with at most $L$ sensitivity, AboveThreshold algorithm is $\epsilon$-DP for every $\epsilon > 0$ \cite{lyu2016understanding}.}
\end{defn}

\begin{defn} \label{def:composition} [Composition] Let a set of randomized algorithms $\mathcal{A}_{1},..., \mathcal{A}_{k}$ that each $\mathcal{A}_{i}$ satisfies $\epsilon_{i}$-DP. 
\begin{itemize}
    \item \textit{Sequential Composition:} Let $\mathcal{A}$ be another randomized mechanism that executes $\mathcal{A}_{1},..., \mathcal{A}_{k}$ with independent randomness for each $\mathcal{A}_{i}$, then $\mathcal{A}$ satisfies ($\sum_{i}\epsilon_{i}$)-DP. 
    \item \textit{Parallel Composition:} Let dataset $\mathbf{X}$ is partitioned depterministically to different subsets $\mathbf{X}_{1},...,\mathbf{X}_{k}$ and executing each $\mathcal{A}_{i}$ with a different disjoint set $\mathbf{X}_{i}$ satisfies $\max_{i}{(\epsilon_{i})}$-DP.
    \item Post-processing a randomized algorithm $\mathcal{A}$ that satisfies ${\epsilon}$-DP does not break or consume any privacy budget. 
\end{itemize}
\end{defn}
\textcolor{black}{Given the composition properties and total $\epsilon$ privacy budget, DP-ANI builds different blocks according to composition properties to achieve a DP-satisfied randomized algorithm $\mathcal{A}$. 
}

\section{Our Metrics and Models}\label{s:metrics}

Protecting personally identifiable information is a crucial step before publishing the output of the queries. This section defines the existing structure and the privacy models we consider for our mobility dataset. 

There are several things to be considered before applying our privacy model on mobility datasets. The privacy-protected aggregated mobility dataset includes both fleet and consumer trajectories. Fleet trajectories may reveal business-related information, such as customer pick up and drop off locations. Similarly, consumer trajectories can reveal daily behaviors of individuals. We solve the problem with two steps: (i) select a perturbation rate adaptively with respect to road network density for each OD of trajectories and (ii) perturb all OD GPS points and match them with new link. With our approach, the output aggregated mobility network is free from privacy issues of individual trajectories. We transform the point-wise GPS trajectories to an ordered series of road network links and enforce privacy on the aggregation of such trajectories. 


Let $D(V,E)$ represent the road network as a weighted digraph, where the set of nodes $V$ correspond to road intersection, set of edges $E$ to roads, and weights that represent link metrics, such as length of the link or traffic volume. A link $\phi \in E$ connects intersections $u$ and $v$ where specific link attributes, such as number of lanes, speed limit, are stored in the link description. We have two set of trajectories: GPS trajectories and link trajectories. Let us define the GPS trajectories and then link trajectories:

{\bf 1) GPS Trajectories}: A sequence of GPS coordinates with $l$ number of samples $ \mathbf{x} \in \mathcal{G} = \{\mathbf{x}_{1}, \mathbf{x}_{2}, ..., \mathbf{x}_{l}\}$ forms a GPS trajectory that reflects the continuous motion of the object. The set of all GPS trajectories are $\Psi$ where $\mathcal{G} \in \Psi$.

{\bf 2) Link Trajectories}: Given the $m$ number of vehicles on the road network, each vehicle travels between origins and destinations using an ordered link path generating a user travel path known as a \textit{micro-graph} $\Phi \in D$. Every link trajectory has $n$ number of links $\phi \in \Phi = \{\phi_{1}, \phi_{2}, ..., \phi_{n}\}$ and $\Phi \subset{E}$. The set of trajectories is the corpus of all link trajectories with $m$ users $\Phi \in \Lambda=\{\Phi_{1}, \Phi_{1}, ..., \Phi_{m}\}$.
A link $\phi_{i,j}$ refers to the $i$th link of user $j$. 
Every link $\phi_{i,j} \in \Phi$ has a set of values: length of the link, travel time, speed, and link counts. The traffic density of the network is represented by aggregating trajectories and is referred to the \textit{ aggregated mobility network}. The raw mobility network is $\vartheta$ and the privacy preserved mobility network is $\Sigma$. The goal of our research is to release the privacy preserved aggregated mobility network from the raw mobility network $\vartheta \rightarrow \Sigma$. 

Link count $\beta$ refers to the number of times a link $\phi$ occurs on the mobility network $\vartheta$. For example, if a link $\phi_{i,j}$ of selected trajectory $\Phi_{j}$ has occurred in the graph $\vartheta$ only once, then 1 is assigned for the link count value for corresponding link. 
The represented link model is built upon the road characteristics. Each link $\phi$ is classified into one of five classes in terms of the capacity and functional role of the road, called a functional class. Arterial roads have lower functional classes, rural streets have higher functional classes. Next, we introduce link matching.

{\bf 3) Link Matching}: GPS coordinates are an estimate of a device's location from satellite broadcast information and are generally enhanced with localized terrestrial information. These locations can be perturbed by localized environments, such as tree cover or urban canyons. As a consequence, GPS locations may not match to a link on the road network. Link matching generates ordered set of road network links describing the user's trajectory considering the road network $D(V,E)$ and GPS points \cite{quddus2007current}. In this paper, we used a link-matching algorithm designed by University of California, Berkeley Smart Cities and Sustainable Mobility Research Group \cite{Macfarlane2021mobile}.

%% file: Noise.tex
\section{Differentially Private Adaptive Noise Injection} \label{s:Noise}

Differential privacy is a probabilistic approach that provides privacy primarily by using injected noise. This noise injection aims to hide individual contributions to the overall statistic while preserving the statistical properties of data in the aggregate level. 
To make origins and destinations of trajectories differentially private at aggregated mobility network, our DP algorithm uses the first approach by injecting planar Laplace noise to the GPS points before generating an aggregated mobility network. The noise level changes the underlying characteristics of the output. The higher noise level leads to stronger privacy but less accurate results. For large and dense datasets, this inaccuracy will be less significant in the overall statistics because the required level of noise to achieve differential privacy is lower \cite{desfontaines2020sok}. Now we explain our noise injection approach.

\begin{figure}[h]
\centering
\includegraphics[width=.45\textwidth]{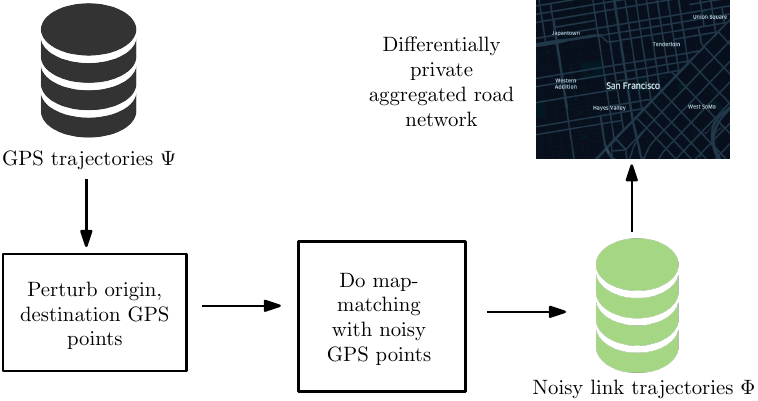}
\caption{Differentially private aggregated mobility network is released using our noise injection model and link matching algorithm. }
\label{f:DP_noise}
\end{figure} 

Fig. \ref{f:DP_noise} shows the flowchart of our differentially-private, aggregated mobility network concept. A user starts from a geographical location called the \emph{origin} and stops at another geographical location called the \emph{destination}. These origin and destination locations are considered sensitive information that must be protected. A \emph{trajectory} consists of the sampled locations between origin and destination. The link-matching algorithm infers the ordered set of links using GPS location data and the previously matched link path from the $D(V,E)$ network for each user's GPS points. The network $D(V,E)$ often constrains the link selection process. 
Since we release the aggregated mobility network, the privacy issue is reduced to the link set that has a low density of links and are connected to origins and destination nodes.

To provide privacy to the aggregate mobility datasets, we use a differentially-private, adaptive noise injection (DP-ANI) model, using planar Laplace noise. The origin and destination GPS points are obfuscated based on the network density, and noisy GPS points are matched with a new link using the link-matching algorithm. 
The two key parameters used for noise sampling are $\epsilon$ and $R$. While $\epsilon$ is responsible for the noise level, $R$ is the distance parameter for moving the center of the noise in the geospatial domain. Output of the noise function is a new randomized GPS location in the same space. The noise function is a bounded probability distribution on polar coordinate systems.

 

\subsection{Noise Sampling}\label{s:noise_sampling}

One state of the art location-hiding approach to inject noise to the GPS dataset is called geo-indistinguishability \cite{andres2013geo}. The mechanism adds planar Laplace noise to the GPS point $x_0$ within an area of $r$. The output of geo-indistinguishability is a perturbed GPS location whose privacy level is $\epsilon r$. The radius $r$ is the desired distance to provide privacy protection. 

Geo-indistinguishability provides a guarantee that the probability of the exact GPS point presence decreases exponentially with the radius $r$ for given location $\mathbf{x_0}$. This noise function is a linear distribution for problems in 1-D space. It is a 2-D surface for a geospatial problem. The probability density function (PDF) of such noise function for any point $\mathbf{x}\in \mathcal{R}^2$ is:

\begin{equation}
    D_{\epsilon}(\mathbf{x_0})(\mathbf{x})=\frac{\epsilon^2}{2\pi}e^{-\epsilon d(\mathbf{x_0},\mathbf{x})}
\end{equation}
where $\frac{\epsilon^2}{2\pi}$ is a normalization factor. This PDF function is a planar Laplacian distribution that is sampled in polar coordinates instead of Cartesian coordinates $D_{\epsilon}(r, \theta)=\frac{\epsilon^2}{2\pi}re^{-\epsilon r}$. A point in polar coordinates $(r, \theta)$, where $r$ is the distance of $\mathbf{x}$ from $\mathbf{x_0}$ and $\theta$ is the angle, is randomly drawn. Since $r$ generally adds a small perturbation to the GPS point, in this work, the $r$ value is scaled with given $R$ radius value from the input in order to provide adaptive noise structure. By scaling up the obfuscation parameter $r$ we have a larger noise level that moves the GPS point further away in euclidean distance. The noise sampling approach involves the following steps:

\begin{itemize}
    \item Draw $\theta$ uniformly in $[ 0, 2\pi)$,
    \item Draw $p$  uniformly in $[0,1)$,
    \item Find $\Gamma=C^{-1}_{\epsilon}(p)$,
    \item Set $r=\Gamma*R$ for larger noise with given input radius $R$
    \item Finally, calculate $\mathbf{z}=\mathbf{x}+ [r\cos(\theta), r\sin(\theta)]$,
\end{itemize}
where $C^{-1}_{\epsilon}(p)=-\frac{1}{\epsilon}(W_{-1}(\frac{p-1}{\epsilon})+1)$ is the inverse cumulative distribution function of $r$, and $W_{-1}$ is Lambert $W$ function (the ${-}1$ branch). 

\textcolor{black}{The original implementation of geo-indistinguishability achieves privacy through fixed bounding range $R$. However, our approach provides geo-indistinguishability by adding controlled noise $L(\epsilon, R)$ to the origin and destination GPS points $x_i$ within a certain range $R$ in order to mask the actual locations using density-based private noise range selection method $R$. Selecting the same threshold from all noise function would allow adversaries to access private information through reverse engineering. Our work also privatizes the range value $R$ with a different DP mechanism. }

\begin{figure}[t]
\centering
\includegraphics[width=.45\textwidth]{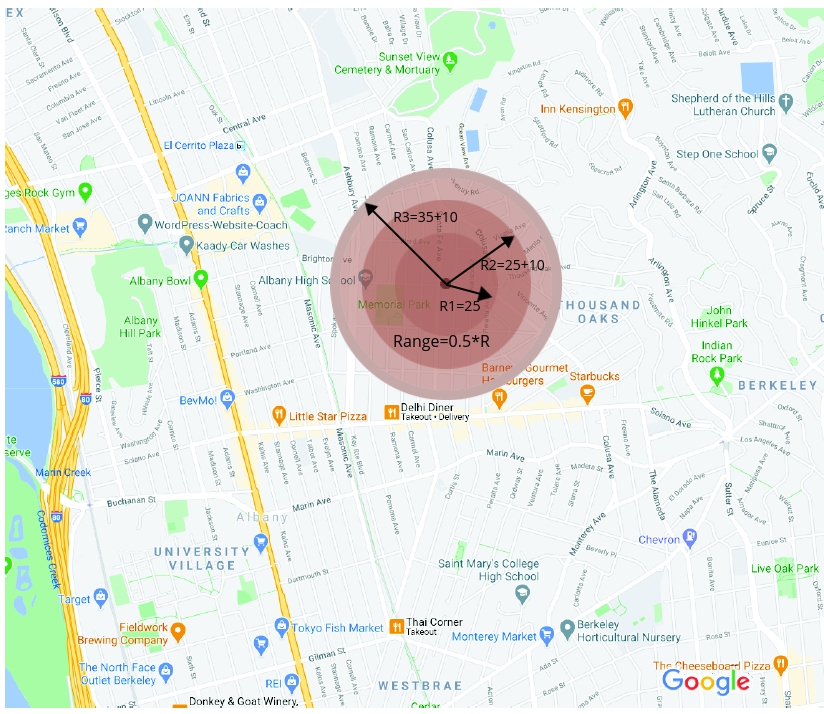}
\caption{Buffer range for determining link density}
\label{f:Range_buffer}
\end{figure}

\subsection{Private Bounding Range Selection} \label{s:ani}
\textcolor{black}{
This section explains determining the bounding range and selecting the noisy link for the link-matching approach using planar Laplace noise introduced in the previous section. Randomly injecting noise without considering the network's density would not achieve the desired privacy level consistently. Some GPS points with few links nearby would match the noisy GPS point with the same link after the noise injection. To overcome this problem, the noise level is selected adaptively with respect to the link density of network $D(V,E)$.}  

\textcolor{black}{The adaptive bounding range selection may also reveal some information about the user's location. Therefore, we employed a private bounding range selection algorithm using the sparse vector technique \cite{lyu2016understanding} to find the bounding range $R$ privately. We were inspired to use the private parameter selection method for geospatial domains by \cite{ghazi2022private}. }

\begin{algorithm}[h]                  
\caption{Private Bounding Range Selection}      
\label{a:adaptive_noise}                         
\begin{algorithmic}[1]                   
\STATE \textbf{\emph{Input}} privacy budget $\epsilon_{radius}$, threshold $\tau$, initial bounding circle $z_{init}$, maximum bounding circle $R_{max}$, network $D(V,E)$, and number of iterations $k = R_{max}/Z$, link functional class $fc$.
\STATE $z \leftarrow z_{init}$
\FOR{$\ell=1,...,k$}
    \STATE $N_{\ell} \leftarrow$ Number of links within $z$ at network $D(V,E)$
    \STATE $z$+=10 meters
\ENDFOR
\STATE $\ell^{*} \leftarrow AboveThreshold(N_{1},...,N_{k}, \tau, \epsilon_{radius})$
\STATE $R \leftarrow z_{init}*\ell^{*}$   
\STATE $BoindingSetFC$ $\leftarrow$ links within $R$ with functional class $fc$
\RETURN $R$ and $BoindingSetFC$
\end{algorithmic}
\end{algorithm}

Our approach generates noisy link-based trajectories while keeping the noisy trajectories close enough to the original trajectories to have similar traffic characteristics at aggregated mobility network. As we mentioned above, every link has functional class information, and the density-based noise function should move the GPS point to a place that matches with the same functional class.  

\begin{defn} \label{def:NoiseFunction} [Density Function] Given the $\epsilon$ value, bounding range $R$ of the noise function $L(\epsilon, R)$ is selected privately using SVT differential privacy mechanism using the function $f(\theta)$ where $\theta$ is the network density in terms of the number of links.
\end{defn}

\textcolor{black}{
Our noise model perturbs every origin and destination of trajectories and applies link-matching to the noisy trajectories. Then, the aggregated mobility network is obtained with a DP guarantee.} 


\begin{algorithm}                      
\caption{Noisy link matching algorithm}      
\label{a:Noisy_link_matching}                         
\begin{algorithmic}[1]                   
\STATE \textbf{\emph{Input}} $\Lambda, {\Psi}$
\STATE \textbf{\emph{Input}} $\epsilon_{laplace}$ for planar Laplace noise function
\FOR{${\mathcal{G} } \in \Psi $}
    \FOR{$x \in {\mathcal{G}}$} 
        \IF{For OD GPS points}
        \STATE Select $R$ and $BufferSetFC$ using Algorithm \ref{a:adaptive_noise}
        \STATE Inject adaptive noise to the GPS point $x$ using $L(\epsilon, R)$
        \STATE Select closest link from $BufferSetFC$
        \ELSE
        \STATE Select closest nodes from $V$
        \ENDIF
    \ENDFOR
    \STATE Find candidate paths between closest nodes list
    \STATE Connect the candidate paths
    \STATE Build the noisy link-matched trajectory with connected links 
\ENDFOR
\RETURN The noisy link trajectory network $\Sigma$
\end{algorithmic}
\end{algorithm}

\textcolor{black}{ The straightforward method would select a bounding range $R$ considering the worst-case scenario given the whole trajectory dataset. However, this would lead to a poor utility at aggregated mobility network because the larger bounding range $R$ takes the center of the noise function to a far distance, resulting in higher perturbation on GPS location. Instead, we developed an algorithm to select the bounding range $R$ privately in Algorithm \ref{a:adaptive_noise}.}

\textcolor{black}{The method first adjusts the noise level using the link network density around the GPS point (see Fig. \ref{f:Range_buffer}). To do so, starting from an initial radius $Z=Z_{init}$, the method increases the radius $k$ iterations until reaching the maximum bounding range $R_{max}$. The $R_{max}$ would be the area that covers the whole geospatial region of the dataset. For each iteration, $N_{\ell}$ contains the number of links within the bounding range $Z$ at network $D(V,E)$. In the second phase, our algorithm selects an index $\ell^*$ given the threshold $\tau$ using \textit{AboveThreshold} algorithm, an SVT algorithm described in \cite{dwork2006differential}. The private bounding range $R$ is acquired with respect to $\ell^*$ and corresponding road links within this bounding range from the same functional class links are stored in $BoindingSetFC$. Please refer to Algorithm \ref{a:adaptive_noise} for the full pseudo-code. }

\textcolor{black}{Once the bounding range $R$ is selected, our algorithm applies noise injection  for the corresponding GPS point using $L(\epsilon, R)$. After the noisy GPS point is returned, the closer link to the GPS point in the set of the same functional class links is selected. Once all the links are found from the GPS points, the candidate paths are selected, and the noisy link trajectory is built using the candidate paths (see Algorithm \ref{a:Noisy_link_matching}). Finally, the noisy link-based trajectories generated with our approach form a mobility network that ensures the privacy of link-level OD locations. }

\subsection{Differential Privacy Analysis}

\textcolor{black}{
This paper provides origin-destination privacy to aggregated mobility networks. For each trajectory, there are two sensitive locations, origin and destination. While two DP mechanisms are used in this paper, bounding range selection and noise injection, the sensitivity $\Delta$ is the same for output data. Adding and removing a user can only change the aggregated network visitation rate at most two links for each origin and destination. Therefore, the sensitivity of our DP mechanisms is 2.}

\begin{lemma}
\textcolor{black}{Our bounding range selection algorithm satisfies $\epsilon_{radius}$-DP.}
\end{lemma}

\begin{proof}
\textcolor{black}{Given the sensitivity of two queries $N_{\ell}$, where adding or removing a user location can change the link visitation at most two links by one, and the definition of SVT \cite{lyu2016understanding}, the $AboveThreshold$ algorithm is $\epsilon_{radius}$-DP. Accordingly, the private bounding range selection algorithm is also $\epsilon_{radius}$-DP.}
\end{proof}

\begin{lemma}
\textcolor{black}{Given the sequential and parallel composition properties of DP, the noisy link matching algorithm satisfies 2*($\epsilon_{laplace}$+$\epsilon_{radius}$) DP for the output mobility network. }
\end{lemma}

\begin{proof}
\textcolor{black}{The aggregated mobility network is produced from the $n$ user trajectories. Each trajectory has one origin and one destination GPS point. For each of these GPS points, noisy link matching algorithm selects a private bounding range $R$ and adds planar Laplace noise using the private $R$. Given the sequential composition property, total privacy budget for each trajectory is $\epsilon_{user} = 2*(\epsilon_{laplace}+\epsilon_{radius})$. Since each trajectory accessed only once and uses the same privacy budget $\epsilon_{user}$, the overall privacy budget of output aggregated mobility network is $\epsilon_{user}$ given the parallel composition. In summary, our private aggregate mobility network satisfies $\epsilon_{user}$-DP.}
\end{proof}

%% file: Results_new.tex
\begin{figure}[t]
\centering
\includegraphics[width=.45\textwidth]{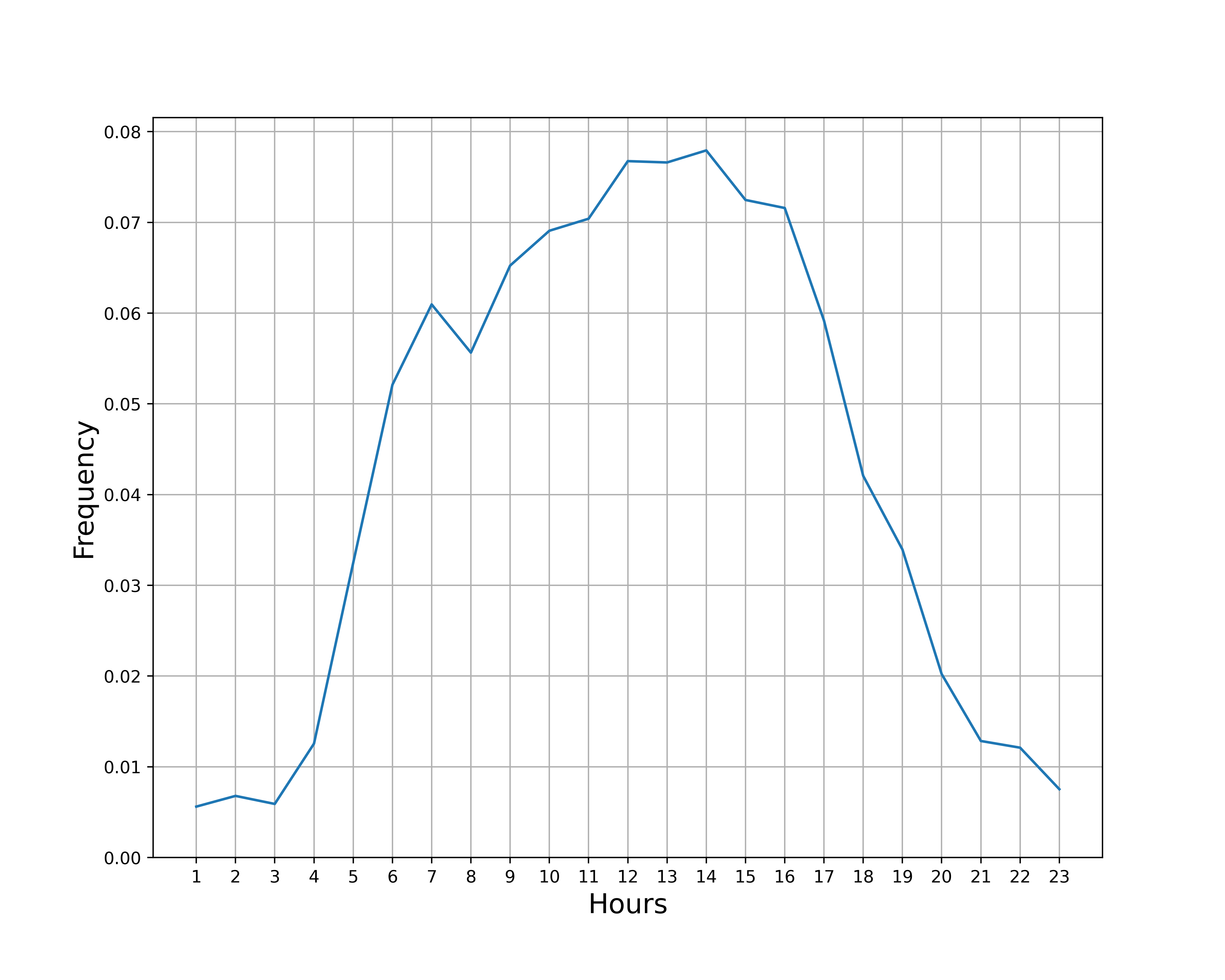}
\caption{OD pair count for a typical week-day} 
\label{f:ODpaircount}
\end{figure}

\section{Experimental Evaluation} \label{s:results}

\subsection{Dataset Description}

This project uses a real-world dataset collected in the San Francisco Bay area in California from January to February 2019. It contains two months of fleet and consumer GPS trajectories with varied sampling rates. To make the computation tractable, we extract a smaller region in the Berkeley area and time period for evaluating our privacy model. Most of the trajectories have sampling rates of less than 1-minute, which improved the ability of the link-matching algorithm to generate realistic link trajectories.

The dataset is created from a variety of location-sharing applications and GPS tracking devices. When the devices are active, location (lat, lon), speed, and heading are collected along with a unique device identifier. While we do not necessarily know that the device was active at the user's actual origin and destination points, we assume that the start and end points of GPS traces represent origins and destinations.


\subsection{Temporal Correlation} \label{s:temporalcorrelation}

Fig. \ref{f:ODpaircount} shows the hourly distribution of the OD pairs in the San Francisco region. A temporal pattern is associated with the time of the day, with distinct rush hours where the number of OD pairs peaks. Hence, any aggregation methods should preserve this pattern if it is to be useful for traffic management applications. For example, aggregating mobility data from morning hours with afternoon non-busy hours would not reflect the real traffic patterns for the morning or afternoon. Therefore, we apply our privacy-preserved aggregation model for each 1 hour period. We show as an example the 1 pm to 2 pm time period. 

\subsection{Comparisons to Alternate Approaches}

We compare the DP-ANI model to several techniques. A straightforward privacy method, similar to the k-anonymity approach in \cite{monreale2010movement}, removes successive links, either from the origin or to the destination, with less than $k$ link counts from the aggregated network. We have chosen $k=2$ for our experiments. We refer to this model as \textit{OD successive remove}. 

The privacy definition of DP-ANI is based on distance, the shortest distance between noisy and the original location. Therefore, we also included an ablation study to have a fair understanding for our DP-ANI model:
\begin{itemize}
    \item \textbf{$DP-ANI_{Fix}$:} This version performs the same privacy mechanism for origin-destination without a private range selection. Here we assumed that the data curator sets a fixed threshold of a number of links for adaptive range selection, which is 25 road links for our network $D(V,E)$. 
    
    \item \textbf{$DP-ANI_{Max}$:} Standard DP methods, mainly for tabular datasets, applies noise injection based on the sensitivity with respect to the worst case scenario. However, geospatial datasets are different, and their sensitivity definition differs from others. The other baseline model applies the worst-case maximum range to all trajectory ODs. Instead of setting the number of links and looking for a range value, we found the maximum range value with more than 25 road links in whole trajectory ODs.  
\end{itemize}

The DP-ANI provides stronger privacy by trading off some utility of the dataset (i.e., answering a subset of finer-granularity queries). We show the performance of the proposed mechanism with respect to the original aggregated model without any privatization method in our experiments as \textit{Original data}.

\subsection{Utility Metrics}

\begin{figure}[h]
\centering
\includegraphics[width=.45\textwidth]{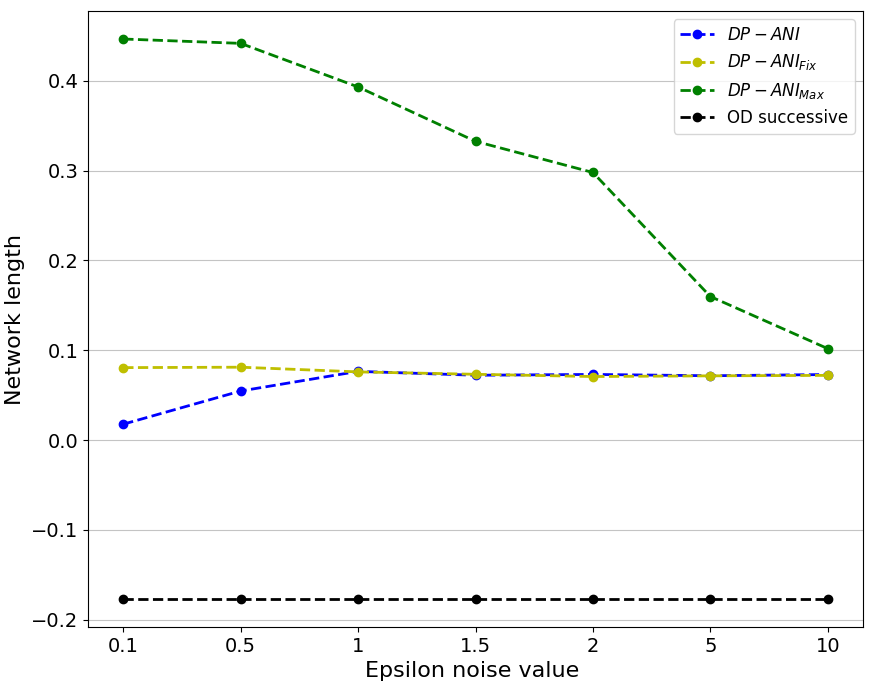}
\caption{Network length for all trajectories with baseline comparisons.}\label{f:NetworkLength}
\end{figure}

We have chosen practical utility metrics commonly used in transportation studies to quantify the efficiency of our privacy mechanism. In this section, we explain the importance of utility metrics. The goal is to have a higher similarity in the utility metrics between the original and privacy-preserved mobility networks, given the same level of privacy protection. In order to present results clearly, all the numerical results are normalized with respect to the original mobility network. 

Spacial density analysis plays a crucial role in understanding human mobility \cite{hasan2013spatiotemporal}. Our first utility metric is the change in aggregated mobility network length. This is our primary utility metric since we aim to acquire a similar mobility network that retains the mobility characteristics. Minimizing the change in the length of aggregated mobility network makes output privatized data more useful.

Since our mechanism provides privacy for user ODs, the second metric we looked at is trajectory level utility: the rate of OD link that moves to another road link after noise injection. The goal is to displace as many OD links as possible.

The main characteristic of aggregated mobility networks is the link visitation rates or link counts. There are several distance metrics to compare the probability distributions. The Wasserstein distance metric is a good way of comparing the count query distributions quantifying the similarities of probability distributions given a metric space. Our count query metric is the number of link visitations named in figures as \textit{Link Counts}.  

The road network classifies the link in different classes, from local streets to highways, into five classes. It is essential to retain the road class distribution in the output network similar to the original aggregated network. We employed the Wasserstein distance on the road link functional class distribution for another aggregated utility metric.

\begin{figure*}
\subfigure[Only the original network]
{\centering
\includegraphics[width=0.3\textwidth]{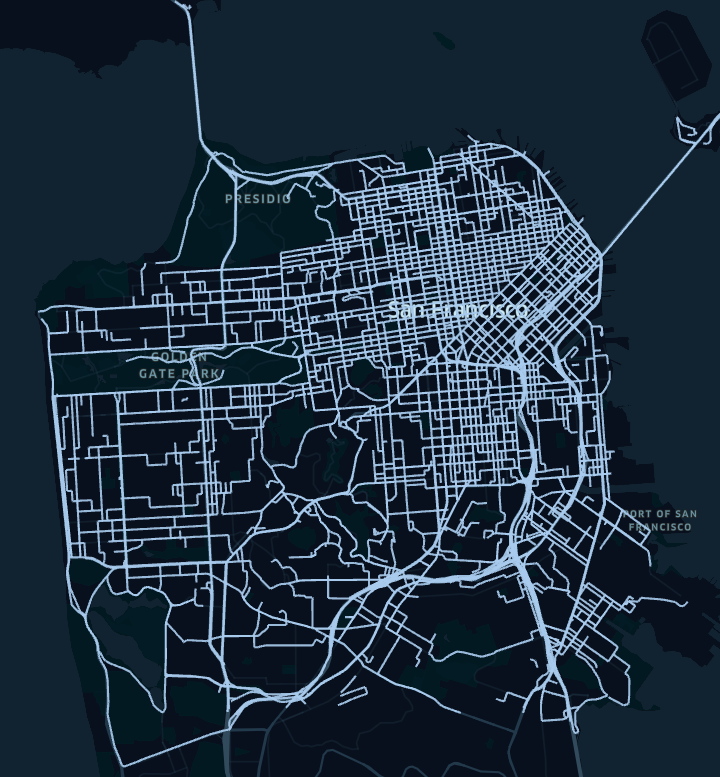}
\label{f:original_network}
}
\subfigure[Original vs the DP-ANI model network]
{\centering
\includegraphics[width=0.3\textwidth]{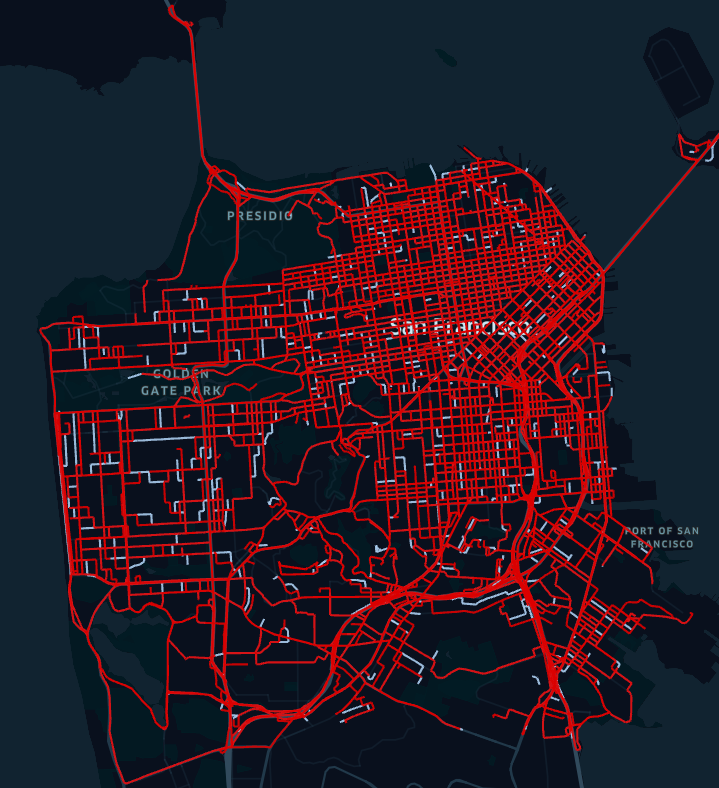}
\label{f:original_vs_ani}
}
\subfigure[Original vs OD-successive link removal network]
{\centering
\includegraphics[width=0.3\textwidth]{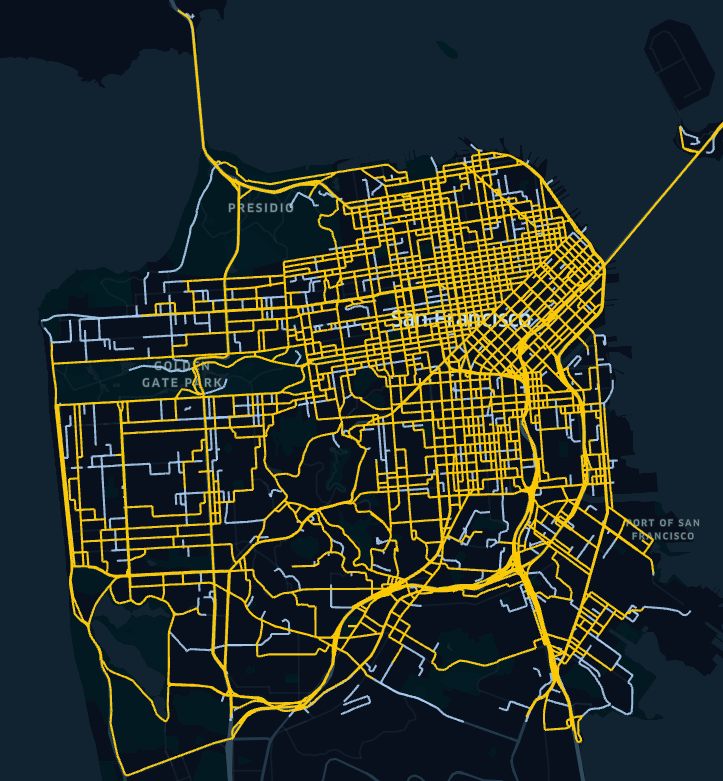}
\label{f:original_vs_od}
}
\centering
\caption{Aggregated trajectory networks for a 1 hour period are represented together on the map. The red link network represents the DP-ANI-AT model for $\epsilon=1$, the yellow represents the OD successive link removing model, and the light blue represents the original model. Fig.-(a) shows the original network only without any overlap. In Fig.-(b) and Fig.-(c), when the DP and OD-successive links overlap with the original network, the links appear as red and yellow, respectively. The network lengths are 195.2, 138.2, and 194.6 miles for networks with the DP-ANI-AT model, OD successive link removal model, original network respectively.}
\label{f:Networks}
\end{figure*}

\subsection{Numerical Results}

The DP-ANI method can extend or shrink the length of the trajectories. As such, fluctuations in the experimental results concerning the different levels of $\epsilon$ values are expected. For lower (or higher) $\epsilon$ values, we have higher (or lower) noise levels. A higher noise level will perturb more links. When we apply the DP-ANI method to the mobility dataset, the number of privatized single-link-count links vary depending on the privacy level $\epsilon$. We use a range of $\epsilon$ values between $0.1$ and $10$ to evaluate the performance of the DP-ANI. The chosen $\epsilon$ values are selected to reflect the lower and upper limits of the impact of the DP-ANI perturbation mechanism. We present our results with the same aggregation concept described in \ref{s:temporalcorrelation} and the same set of $\epsilon$ values for all the experiments.


The DP-ANI may affect the active network length as the noise may adjust the trajectory to a position that requires more maneuvers or positions a start location behind the original origin location or similarly ahead of the original destination. Fig. \ref{f:NetworkLength} shows the network lengths of different privacy models with the set of $\epsilon$ values in terms of the divergence from the original aggregated network. The blue line shows the performance of our adaptive threshold DP-ANI model. 1.0 in the y-axis is the reference point for the original network. The DP-ANI model has the slightest change from the original network compared to other variants with less than $10\%$. While $DP-ANI_{Max}$ is the second-best-model, $DP-ANI_{Max}$ and \textit{OD successive} yielded very high divergence from the original network. Note that since \textit{OD successive} do not depend on the $\epsilon$ parameter, their resulting plots are flat across the horizontal axis.

Next, we visualize the aggregated networks to see the differences between the DP-ANI-AT model with $\epsilon=0.1$ and other comparative models in Fig. \ref{f:Networks}. \textit{OD-successive}, a k-anonymity based approach, is another formal privacy protection model. For this reason, we compare our model with \textit{OD-successive} and original network models. To preserve the utility of the privacy-preserving data, we would like the aggregated trajectory network to overlap with the original network except for origin/destination points/segments. In Fig. \ref{f:original_network}, we only present the original aggregated network to reflect the actual coverage area. While the  DP-ANI network shows a high similarity with the original network (Fig. \ref{f:original_vs_ani}), there is a distinct difference between \textit{OD-successive} and original networks shown in Fig. \ref{f:original_vs_od}, primarily in the outer areas of the San Francisco region.

\begin{figure}[h]
\centering
\includegraphics[width=.45\textwidth]{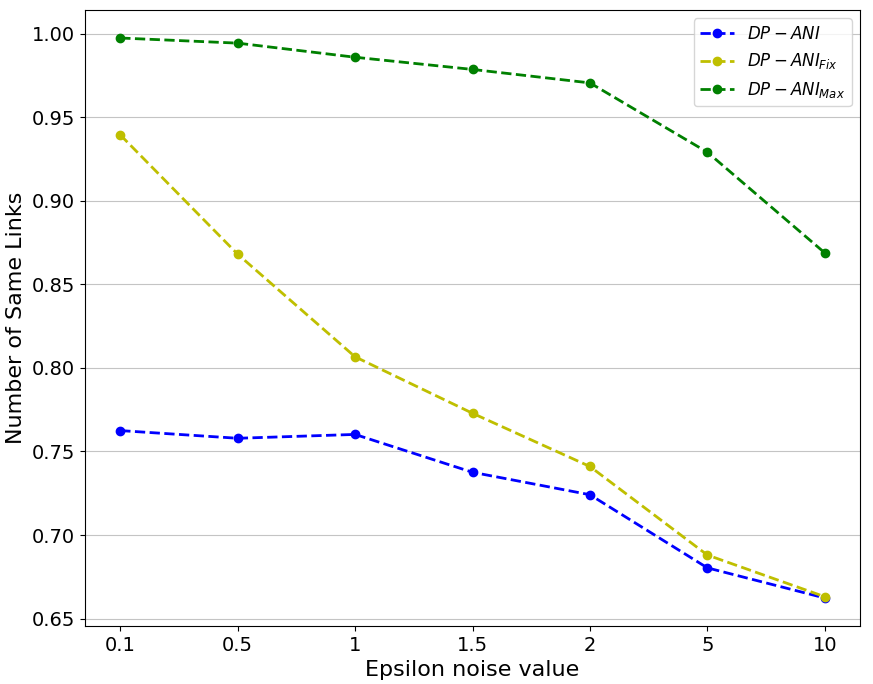}
\caption{Comparison of different $\epsilon$ values and the change of percentage of ODs. }\label{f:SameLinks}
\end{figure}

Next, we identify the ratio of unchanged OD links after the noise injection and aggregation steps. We only look at OD link displacements since we consider the link as not having any privacy concerns if it is not an origin or destination for any trajectory. Depending on the noise level and the density of the road network $D(V,E)$, the links can be matched with the same link after the noise injection. The value of $\epsilon$ defines the level of privacy. To quantify the privacy of our method, we inspect the privatized link ratio over the OD links with respect to different $\epsilon$ values. Fig. \ref{f:SameLinks} shows the ratio of unchanged OD links for all ODs. The goal of the DP-ANI model is to move OD links. Therefore the output is expected to have higher ratios for the lower level of $\epsilon$. The highest level of the privatized link ratio is observed with the lowest $\epsilon=0.1$ with an average of $76\%$. $DP-ANI_{Max}$ and $DP-ANI_{Fix}$ achieve higher ratios compared to DP-ANI. However, they both have drawbacks: $DP-ANI_{Max}$ results in lower utility and $DP-ANI_{Fix}$ provides a lower privacy guarantee with a fixed range value that could leak more information due to less randomization. The lowest $\epsilon$ produces the fewest number of single-link-count OD-links. 

\begin{figure}[h]
\centering
\includegraphics[width=.45\textwidth]{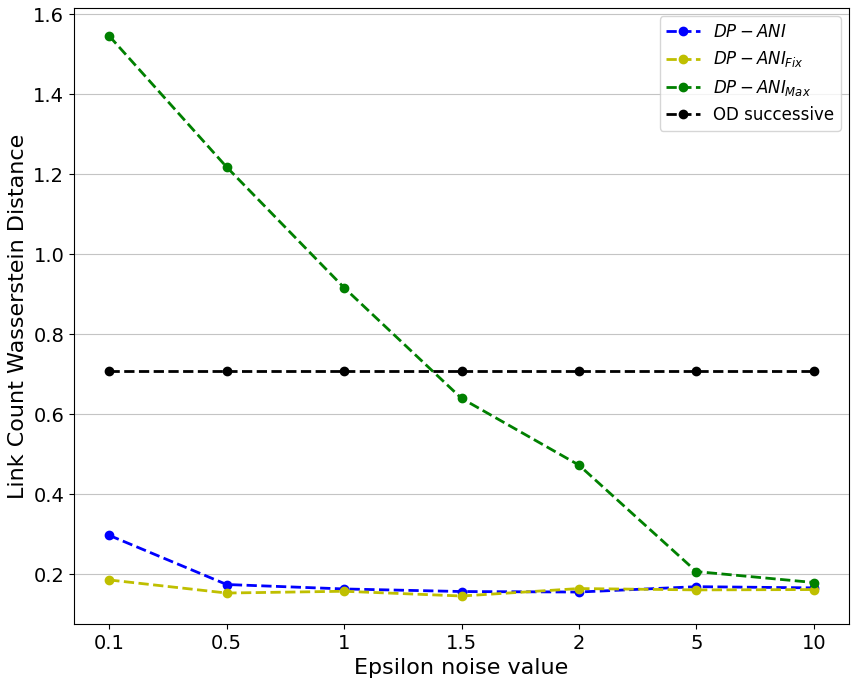}
\caption{Wasserstein distance of link counts}\label{f:LinkCountWD}
\end{figure}

\begin{figure}[h]
\centering
\includegraphics[width=.45\textwidth]{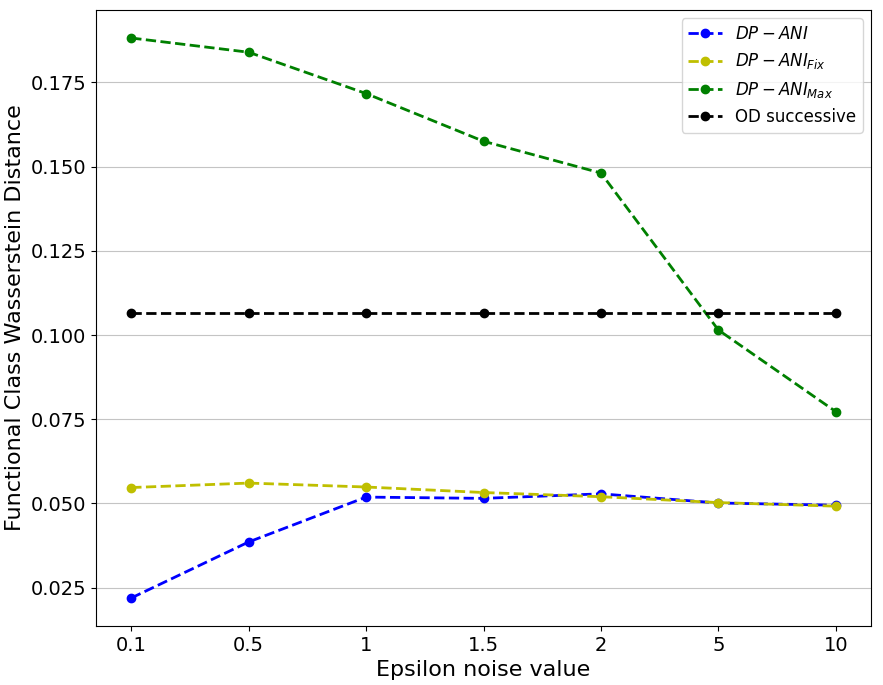}
\caption{Wasserstein distance of link functional classes}\label{f:FuncClassWD}
\end{figure}

Finally, we evaluate the performance of DP-ANI model with different variants and \textit{OD-successive} model on the Wasserstein distributional similarity metric in Fig. \ref{f:LinkCountWD} and Fig. \ref{f:FuncClassWD}. The lower value indicated higher similarity. For all the $\epsilon$  values, our method is more similar to the original aggregated mobility network at link count distribution and the functional class distribution. Increasing the $\epsilon$ value yields higher similarity in distributions for other variants DP-ANI-Max and DP-ANI-Fixed with a lower privacy guarantee.

%% file: Discussion.tex
\section{Discussion} \label{s:discuss}
Privacy of aggregated mobility trajectories serves as a useful first step towards unlocking siloed trajectory data. Because geospatially constrained and temporally correlated mobility datasets have unique characteristics, it is difficult to preserve the privacy of individual trajectories. This paper describes a method generating an aggregated mobility representation of the original trajectory set using DP-based adaptive noise injection model. In this section, we summarize several limitations and discuss open research questions.

\subsection{Limitations}

The first limitation of our study is aggregation scale. The generated aggregate mobility network aims to reflect the general patterns of human mobility, however, due to sparsity of most of the datasets that we used, the realistic travel patterns are not evident in the aggregated dataset as shown in the earlier Figure \ref{f:ODpaircount}, where the approximately 1200 trajectories do not show the expected rush hour travel pattern. The sparsity of the dataset required high perturbation in order to privatize the dataset. In order to overcome this problem, aggregating three or more days of trajectories in one hour period can preserve the utility of the dataset, but this characteristic is highly dependent on the dataset. 

Inherently perturbing origins and destinations can affect the meaning of the object. Although DP-ANI maintains the same functional class of the OD links, it can cause a catastrophic change in the individual trajectory that will affect the aggregated output. For example, when we perturb a link to the other side of the highway, or to a one way road, that moved link must connect with the rest of the trajectory. This may result in much longer routes and if the composition of the dataset involves many of these type of perturbations, this will significantly change the total travel distances of trajectories. 

Finally, biases associated with fleet trajectories is a third consideration. Fleet trajectories should be privatized as patterns related to the fleet's business-model, however, this approach likely does not solve the fleet privacy problem, as a fleet's patterns may still be evident in the aggregated mobility model. Ideally, it is better to work with an evenly distributed fleet and consumer trajectories in the aggregated output to reflect the true travel patterns. Depending on the original dataset, removing fleet trajectories and working only with consumer trajectories may result in a very sparse dataset. This study used both fleet and consumer trajectories. 

\subsection{Open Research Directions}

The DP-ANI model achieves an ideal level of privacy by preserving the privacy of more than 75\% of the privacy concerned ODs on three aggregation models at around $\epsilon=0.1$ for the experimented San Francisco area. Regarding to the privacy level, it could be interesting to test DP-ANI on a larger geographical region in order to define a proper $\epsilon$ range. 

Future research will focus on generalizing the DP-ANI model to a broader range of aggregation concepts with different levels of granularities, such as zone level or OD levels, in order to assess confidentiality and applicability of the current approach. In addition, the computational cost of data processing on large datasets will likely require novel computational methods. A recent trend is to minimize centralized data collection and privatize the data at local or regional controller using real-time localized DP models\cite{wang2020comprehensive}. Forming trajectories with perturbed or privatized local GPS locations may break the coherence of the trajectory and the usefulness of the trajectory. As such, investigation into how this localized privacy model affects the aggregation and privatization methods would be important. 


%% file: Conclusion.tex
\section{Conclusion}\label{s:conclusion}

This paper presents a differentially-private, adaptive noise injection model for aggregated trajectory networks that protect individual origin-destination locations. This method injects planar Laplace noise to the individual origin-destination GPS points by considering the density of the localized road networks. The actual perturbation distance for each GPS point is adjusted by considering the localized link density and this selection is performed privately with the Adaptive Thresholding method. After injecting noise into the GPS points, the location is matched to new network links, and a new origin-destination privatized trajectory is generated and integrated into the aggregated mobility road link network. We evaluated our differentially private mechanism for a variety of variants and a k-anonymity-based privacy model.

This project uses an aggregation concept that generates a privatized, aggregated mobility network for a specific regional dataset in San Francisco, CA. Future work will include extending this investigation to different types of mobility data aggregation models while also addressing the computational efficiency for large-scale datasets. 


\section{Acknowledgements}

The authors are grateful to Thomas Strohmer, Girish Kumar, Reinhard Gentz and Ziyue Dong for their many helpful comments and feedback throughout the project. We also would like to thank Nikhil Ravi and Raksha Ramakrishna for their proofreading. This work was supported by Contractor Supported Research (CSR) funds provided by Lawrence Berkeley National Laboratory, operated for the U.S. Department of Energy under Contract No. DE-AC02-05CH11231, and Department of Transportation - University Transportation Centers, "Center for Transportation, Environment, and Community Health (CTECH)".